% determine hypertex mode
\newif\iflanl
\openin 1 lanlmac
\ifeof 1 \lanlfalse \else \lanltrue \fi
\closein 1
\iflanl
    \input lanlmac
\else
    \message{[lanlmac not found - use harvmac instead}
    \input harvmac
\fi
\newif\ifhypertex
\ifx\hyperdef\UnDeFiNeD
    \hypertexfalse
    \message{[HYPERTEX MODE OFF}
    \def\hyperref#1#2#3#4{#4}
    \def\hyperdef#1#2#3#4{#4}
    \def\hypernoname{}
    \def\e@tf@ur#1{}
    \def\eprt#1{{\tt #1}}
    \def\CERN{\address{CERN, CH--1211 Geneva 23, Switzerland}}
    \def\wl{W.\ Lerche}
\else
    \hypertextrue
    \message{[HYPERTEX MODE ON}
%hypertex links to xxx.lanl.gov:
%  \def\hth/#1#2#3#4#5#6#7{\special{html:<a
%   href="<¬http://xxx.lanl.gov/abs/hep-th/#1#2#3#4#5#6#7">}
%  {\tt hep-th/#1#2#3#4#5#6#7}\special{html:</a>}}
\def\eprt#1{{\tt
#1}}
\def\CERN{\address{

Theory Division, CERN, Geneva, Switzerland}}
\def\wl{

 W.\ Lerche}
\fi
%%%%%%%%%%%%%%%%%%%%%%% %%%%%%%%%%%%%%%%%%%%%%%
\newif\ifdraft

\noblackbox
\catcode`\@=11
\newif\iffrontpage
%%%%%%%%%%%%%%%%%%% %%%%%%%%%%%%%%%%%%%%%%%%%%%%%%%%%%%%%%%%%%%%%%
%%%%% sizes, offsets etc
%%%%%%%%%%%%%%%%%%% %%%%%%%%%%%%%%%%%%%%%%%%%%%%%%%%%%%%%%%%%%%%%%
\ifx\answ\bigans
\def\titleft{\titla}
\magnification=1200\baselineskip=14pt plus 2pt minus 1pt
%
%%%%% unreduced mode: %%%%
%\voffset=0.35truein\hoffset=0.250truein
\advance\hoffset by-0.075truein
\advance\voffset by1.truecm
\hsize=6.15truein\vsize=600.truept\hsbody=\hsize\hstitle=\hsize
\else\let\lr=L
\def\titleft{\titla}
\magnification=1000\baselineskip=14pt plus 2pt minus 1pt
%
%%%%% reduced mode: %%%%%%%
\hoffset=-0.75truein\voffset=-.0truein
%?\hoffset=-.25truein\voffset=-.0truein
\vsize=6.5truein
\hstitle=8.truein\hsbody=4.75truein
\fullhsize=10truein\hsize=\hsbody
\fi
\parskip=4pt plus 15pt minus 1pt
%
%%%%%%%%%%%%%%%%%%% %%%%%%%%%%%%%%%%%%%%%%%%%%%%%%%%%%%%%%%%%%%%%%
%%%%%  fonts
%%%%%%%%%%%%%%%%%%% %%%%%%%%%%%%%%%%%%%%%%%%%%%%%%%%%%%%%%%%%%%%%%
\font\bigit=cmti10 scaled \magstep1

\font\titla=cmr10 scaled\magstep3
\font\tenmss=cmss10
\font\absmss=cmss10 scaled\magstep1

\newfam\mssfam
\font\footrm=cmr8  \font\footrms=cmr5
\font\footrmss=cmr5   \font\footi=cmmi8
\font\footis=cmmi5   \font\footiss=cmmi5
\font\footsy=cmsy8   \font\footsys=cmsy5
\font\footsyss=cmsy5   \font\footbf=cmbx8
\font\footmss=cmss8
\def\footfont{\def\rm{\fam0\footrm}
\textfont0=\footrm \scriptfont0=\footrms
\scriptscriptfont0=\footrmss
\textfont1=\footi \scriptfont1=\footis
\scriptscriptfont1=\footiss
\textfont2=\footsy \scriptfont2=\footsys
\scriptscriptfont2=\footsyss
\textfont\itfam=\footi \def\it{\fam\itfam\footi}
\textfont\mssfam=\footmss \def\mss{\fam\mssfam\footmss}
\textfont\bffam=\footbf \def\bf{\fam\bffam\footbf} \rm}
\def\tenpoint{\def\rm{\fam0\tenrm}
\textfont0=\tenrm \scriptfont0=\sevenrm
\scriptscriptfont0=\fiverm
\textfont1=\teni  \scriptfont1=\seveni
\scriptscriptfont1=\fivei
\textfont2=\tensy \scriptfont2=\sevensy
\scriptscriptfont2=\fivesy
\textfont\itfam=\tenit \def\it{\fam\itfam\tenit}
\textfont\mssfam=\tenmss \def\mss{\fam\mssfam\tenmss}
\textfont\bffam=\tenbf \def\bf{\fam\bffam\tenbf} \rm}
\ifx\answ\bigans\def\abstractfont{\tenpoint}\else
\def\abstractfont{\def\rm{\fam0\absrm}
\textfont0=\absrm \scriptfont0=\absrms
\scriptscriptfont0=\absrmss
\textfont1=\absi \scriptfont1=\absis
\scriptscriptfont1=\absiss
\textfont2=\abssy \scriptfont2=\abssys
\scriptscriptfont2=\abssyss
\textfont\itfam=\bigit \def\it{\fam\itfam\bigit}
\textfont\mssfam=\absmss \def\mss{\fam\mssfam\absmss}
\textfont\bffam=\absbf \def\bf{\fam\bffam\absbf}\rm}\fi
%
%%%%%%%%%%%%%%%%%%%%%%%%%%%%% %%%%%%%%%%%%%%%%%%%%%%%%%%%%%%%%%
%%%%% footnotes   (adapted from PHYZZX, no hypertext yet)
%%%%%%%%%%%%%%%%%%%%%%%%%%%%% %%%%%%%%%%%%%%%%%%%%%%%%%%%%%%%%%
\def\f@@t{\baselineskip10pt\lineskip0pt\lineskiplimit0pt
\bgroup\aftergroup\@foot\let\next}
\setbox\strutbox=\hbox{\vrule height 8.pt depth 3.5pt width\z@}
\def\vfootnote#1{\insert\footins\bgroup
\baselineskip10pt\footfont
\interlinepenalty=\interfootnotelinepenalty
\floatingpenalty=20000
\splittopskip=\ht\strutbox \boxmaxdepth=\dp\strutbox
\leftskip=24pt \rightskip=\z@skip
\parindent=12pt \parfillskip=0pt plus 1fil
\spaceskip=\z@skip \xspaceskip=\z@skip
\Textindent{$#1$}\footstrut\futurelet\next\fo@t}
\def\Textindent#1{\noindent\llap{#1\enspace}\ignorespaces}
\def\foot{\global\advance\ftno by1%
\attach{\hyperref{}{footnote}{\the\ftno}{\footsymbolgen}}%
\vfootnote{\hyperdef\hypernoname{footnote}{\the\ftno}{\footsymbol}}}%
%   this is for custom footnote marks:
\def\footnote#1{\global\advance\ftno by1%
\attach{\hyperref{}{footnote}{\the\ftno}{#1}}%
\vfootnote{\hyperdef\hypernoname{footnote}{\the\ftno}{#1}}}%
\newcount\lastf@@t           \lastf@@t=-1
\newcount\footsymbolcount    \footsymbolcount=0
\global\newcount\ftno \global\ftno=0
\def\footsymbolgen{\relax\footsym
\global\lastf@@t=\pageno\footsymbol}
\def\footsym{\ifnum\footsymbolcount<0
\global\footsymbolcount=0\fi
{\iffrontpage \else \advance\lastf@@t by 1 \fi
\ifnum\lastf@@t<\pageno \global\footsymbolcount=0
\else \global\advance\footsymbolcount by 1 \fi }
\ifcase\footsymbolcount
\fd@f\dagger\or \fd@f\diamond\or \fd@f\ddagger\or
\fd@f\natural\or \fd@f\ast\or \fd@f\bullet\or
\fd@f\star\or \fd@f\nabla\else \fd@f\dagger
\global\footsymbolcount=0 \fi }
\def\fd@f#1{\xdef\footsymbol{#1}}
\def\space@ver#1{\let\@sf=\empty \ifmmode #1\else \ifhmode
\edef\@sf{\spacefactor=\the\spacefactor}
\unskip${}#1$\relax\fi\fi}
\def\attach#1{\space@ver{\strut^{\mkern 2mu #1}}\@sf}
%
%%%%%%%%%%%%%%%%%%% %%%%%%%%%%%%%%%%%%%%%%%%%%%%%%%%%%%%%%%%%%%%%%
%%%%% References
%%%%%%%%%%%%%%%%%%% %%%%%%%%%%%%%%%%%%%%%%%%%%%%%%%%%%%%%%%%%%%%%%
\newif\ifnref
\def\rrr#1#2{\relax\ifnref\nref#1{#2}\else\ref#1{#2}\fi}
\def\ldf#1#2{\begingroup\obeylines
\gdef#1{\rrr{#1}{#2}}\endgroup\unskip}
\def\nrf#1{\nreftrue{#1}\nreffalse}
\def\doubref#1#2{\refs{{#1},{#2}}}
\def\multref#1#2#3{\nrf{#1#2#3}\refs{#1{--}#3}}
\nreffalse
\def\refout{\listrefs}

\def\lref{\ldf}

%%%%%%%%%%%%%%%%%%% %%%%%%%%%%%%%%%%%%%%%%%%%%%%%%%%%%%%%%%%%%%%%%
%%%%%%% eq numbering
%%%%%%%%%%%%%%%%%%% %%%%%%%%%%%%%%%%%%%%%%%%%%%%%%%%%%%%%%%%%%%%%%
\def\eqn#1{\xdef #1{(\noexpand\hyperref{}%
{equation}{\secsym\the\meqno}%
{\secsym\the\meqno})}\eqno(\hyperdef\hypernoname{equation}%
{\secsym\the\meqno}{\secsym\the\meqno})\eqlabeL#1%
\writedef{#1\leftbracket#1}\global\advance\meqno by1}
\def\eqnalign#1{\xdef #1{\noexpand\hyperref{}{equation}%
{\secsym\the\meqno}{(\secsym\the\meqno)}}%
\writedef{#1\leftbracket#1}%
\hyperdef\hypernoname{equation}%
{\secsym\the\meqno}{\e@tf@ur#1}\eqlabeL{#1}%
\global\advance\meqno by1}
%old:
\def\eqnalign#1{\xdef #1{(\secsym\the\meqno)}
\writedef{#1\leftbracket#1}%
\global\advance\meqno by1 #1\eqlabeL{#1}}
%
%%%%%%%%%%%%%%%%%%% %%%%%%%%%%%%%%%%%%%%%%%%%%%%%%%%%%%%%%%%%%%%%%
%%%%%%  macros for titlepage, marginnotes, etc
%%%%%%%%%%%%%%%%%%% %%%%%%%%%%%%%%%%%%%%%%%%%%%%%%%%%%%%%%%%%%%%%%

%
\def\chap#1{\newsec{#1}}
\def\chapter#1{\chap{#1}}
\def\sect#1{\subsec{#1}}
\def\section#1{\sect{#1}}
\def\\{\ifnum\lastpenalty=-10000\relax
\else\hfil\penalty-10000\fi\ignorespaces}
\def\note#1{\leavevmode%
\edef\@@marginsf{\spacefactor=\the\spacefactor\relax}%
\ifdraft\strut\vadjust{%
\hbox to0pt{\hskip\hsize%
\ifx\answ\bigans\hskip.1in\else\hskip .1in\fi%
\vbox to0pt{\vskip-\dp
%\vskip4pt
\strutbox\sevenbf\baselineskip=8pt plus 1pt minus 1pt%
\ifx\answ\bigans\hsize=.7in\else\hsize=.35in\fi%
\tolerance=5000 \hbadness=5000%
\leftskip=0pt \rightskip=0pt \everypar={}%
\raggedright\parskip=0pt \parindent=0pt%
\vskip-\ht\strutbox\noindent\strut#1\par%
\vss}\hss}}\fi\@@marginsf\kern-.01cm}
\def\titlepage{%
\frontpagetrue\nopagenumbers\abstractfont%
\hsize=\hstitle\rightline{\vbox{\baselineskip=10pt%
{\abstractfont\pubnum}}}\pageno=0}
\frontpagefalse
\def\pubnum{}
\def\pdate{\number\month/\number\yearltd}
\def\makefootline{\iffrontpage\vskip .27truein
\line{\the\footline}
%\vskip -.1truein\line{\pdate\hfil}
\vskip -.1truein\leftline{\vbox{\baselineskip=10pt%
{\abstractfont\pdate}}}
\else\vskip.5cm\line{\hss \tenrm $-$ \folio\ $-$ \hss}\fi}
\def\title#1{\vskip .7truecm\titlestyle{\titleft #1}}
\def\titlestyle#1{\par\begingroup \interlinepenalty=9999
\leftskip=0.02\hsize plus 0.23\hsize minus 0.02\hsize
\rightskip=\leftskip \parfillskip=0pt
\hyphenpenalty=9000 \exhyphenpenalty=9000
\tolerance=9999 \pretolerance=9000
\spaceskip=0.333em \xspaceskip=0.5em
\noindent #1\par\endgroup }
\def\autskip{\ifx\answ\bigans\vskip.5truecm\else\vskip.1cm\fi}
\def\author#1{\vskip .7in \centerline{#1}}

\def\address#1{\ifx\answ\bigans\vskip.2truecm
\else\vskip.1cm\fi{\it \centerline{#1}}}
\def\abstract#1{
\vskip .5in\vfil\centerline
{\bf Abstract}\penalty1000
{{\smallskip\ifx\answ\bigans\leftskip 2pc \rightskip 2pc
\else\leftskip 5pc \rightskip 5pc\fi
\noindent\abstractfont \baselineskip=12pt
{#1} \smallskip}}
\penalty-1000}
\def\endpage{\tenpoint\supereject\global\hsize=\hsbody%
\frontpagefalse\footline={\hss\tenrm\folio\hss}}
\def\ack{\goodbreak\vskip2.cm\centerline{{\bf Acknowledgements}}}
%
%

%
%%%%%%%%%%%%%%%%%%%%%%%%%%%%% %%%%%%%%%%%%%%%%%%%%%%%%%%%%%%%%%
\def\bfone{\relax{\rm 1\kern-.35em 1}}
\def\inbar{\vrule height1.5ex width.4pt depth0pt}
\def\IC{\relax\,\hbox{$\inbar\kern-.3em{\mss C}$}}
\def\ID{\relax{\rm I\kern-.18em D}}
\def\IF{\relax{\rm I\kern-.18em F}}
\def\IH{\relax{\rm I\kern-.18em H}}
\def\II{\relax{\rm I\kern-.17em I}}
\def\IN{\relax{\rm I\kern-.18em N}}
\def\IP{\relax{\rm I\kern-.18em P}}
\def\IQ{\relax\,\hbox{$\inbar\kern-.3em{\rm Q}$}}
\def\IR{\relax{\rm I\kern-.18em R}}
\font\cmss=cmss10 \font\cmsss=cmss10 at 7pt
\def\ZZ{\relax\ifmmode\mathchoice
{\hbox{\cmss Z\kern-.4em Z}}{\hbox{\cmss Z\kern-.4em Z}}
{\lower.9pt\hbox{\cmsss Z\kern-.4em Z}}
{\lower1.2pt\hbox{\cmsss Z\kern-.4em Z}}\else{\cmss Z\kern-.4em
Z}\fi}
\def\tx{\theta_x} \def\ty{\theta_y} \def\tz{\theta_z}
 \def\twa{\theta_{w_1}} \def\twb{\theta_{w_2}}

\def\cA{{\cal A}}

\def\cF{{\cal F}}

\def\cL{{\cal L}} \def\cM{{\cal M}}
 
 \def\cQ{{\cal Q}}

\def\nup#1({Nucl.\ Phys.\ $\us {B#1}$\ (}
\def\plt#1({Phys.\ Lett.\ $\us  {#1}$\ (}
\def\cmp#1({Comm.\ Math.\ Phys.\ $\us  {#1}$\ (}
\def\prp#1({Phys.\ Rep.\ $\us  {#1}$\ (}
\def\prl#1({Phys.\ Rev.\ Lett.\ $\us  {#1}$\ (}
\def\prv#1({Phys.\ Rev.\ $\us  {#1}$\ (}
\def\mpl#1({Mod.\ Phys.\ Let.\ $\us  {A#1}$\ (}
\def\ijmp#1({Int.\ J.\ Mod.\ Phys.\ $\us{A#1}$\ (}
\def\tit#1|{{\it #1},\ }
%
%%%%%%%%%%%%%%%%%%%%%%%%%%%%%%%% %%%%%%%%%%%%%%%%%%%%%%%%%%%%%%
%%%%% misc %%%%
%%%%%%%%%%%%%%%%%%%%%%%%%%%%%%%% %%%%%%%%%%%%%%%%%%%%%%%%%%%%%%

%

\def\ni{\noindent}
\def\tilde{\widetilde}
\def\bar{\overline}
\def\us#1{\underline{#1}}

\def\Coe#1.#2.{{#1\over #2}}
\def\coeff#1#2{\relax{\textstyle {#1 \over #2}}\displaystyle}
\def\coe#1.#2.{\relax{\textstyle {#1 \over #2}}\displaystyle}

\def\shalf{\relax{\textstyle {1 \over 2}}\displaystyle}

\def\to{\rightarrow}
\def\notin{\hbox{{$\in$}\kern-.51em\hbox{/}}}

\def\del{\partial}

%%%%%%%%%%%%%%%%%%%%%%%%%%%
\def\eg{{\it e.g.}}
\def\ie{{\it i.e.}}
\def\cf{{\it c.f.}}
%%%%%%%%%%%%%%%%%%%%%%%%%%%
\catcode`\@=12
%%%%%%%%% end macros  %%%%%%% %%%%%%%%%%%%%%%%%%%%%%%%%%%%%%
%%%%%%%%%%%%%%%%%%%%%%%%%%%% %%%%%%%%%%%%%%%%%%%%%%%%%%%%%

% stephan's %%%%%%%%%%
\def\h {{1\over 2}}

\def\ov {\overline}
\def\o {\over}
\def\Li {{\cal L}i}
\def\Uc {{\ov U}}

\def\lf {\left}
\def\ri {\right}

\def\p {\partial}

\def\lf {\Big}
\def\ri {\Big}

%% naming conventions
\def\F{F}

\def\ellgen{{\cA}}
\def\lb#1{^{\{#1\}}}
\def\ot#1{^{\otimes {#1}}}
\def\ow{}

\def\FSTU{{\cal F}}
\def\FTU{{f}}

\def\Fn#1{\FTU\lb{#1}}
\def\kt{^{K3}}
\def\cy{^{CY}}

\def\nihil#1{{\sl #1}}
\def\br{\hfill\break}
\def\np {{ Nucl.\ Phys.} {\bf B}}

%%%%%%%%%%%%%%%%%%%%%%%%%%%% %%%%%%%%%%%%%%%%%%%%%%%%%%%%%
%\input refs
%%%%%%%%%%%%%%%%%%%%%%%%%%%% %%%%%%%%%%%%%%%%%%%%%%%%%%%%%

\lref\Fth{
{C.\ Vafa,
 \nihil{Evidence for F theory,}
 Nucl.\ Phys.\ {\bf B469} (1996) 403-418,
 \eprt{hep-th/9602022}; \br}
{D.\  Morrison and C.\ Vafa,
 \nihil{Compactifications of F theory on Calabi-Yau threefolds~I,}
 Nucl.\  Phys.\ {\bf B473} (1996) 74-92,
 \eprt{hep-th/9602114}; \br
 \nihil{Compactifications of F theory on Calabi-Yau threefolds~II,}
 Nucl.\  Phys.\ {\bf B476} (1996) 437-469,
 \eprt{hep-th/9603161}.}
}

\lref\oneloop{
B.\ de Wit, V.\ Kaplunovsky, J.\ Louis and D.\ L\"ust,
 \nihil{Perturbative couplings of vector multiplets
 in $N=2$ heterotic string vacua,}
 Nucl.\ Phys.\ {\bf B451} (1995) 53-95,
 \eprt{hep-th/9504006};\br
{I.\ Antoniadis, S.\ Ferrara, E.\ Gava, K.\ Narain and T.\ Taylor,
\nihil{Perturbative prepotential and monodromies in $N=2$ heterotic
superstring,}  Nucl.\ Phys.\ {\bf B447} (1995) 35-61,
\eprt{hep-th/9504034}.}}

\lref\SW{N.\ Seiberg and E.\ Witten, \nihil{Electric - magnetic
duality, monopole condensation, and confinement in N=2 supersymmetric
Yang-Mills theory,} \nup426(1994) 19,
\eprt{hep-th/9407087}.}

\lref\kv{S.\ Kachru and C.\ Vafa, \nihil{Exact results for N=2
 compactifications of heterotic strings,}
 Nucl.\  Phys.\ {\bf B450} (1995) 69-89,
 \eprt{hep-th/9505105}.
}

\lref\KLM{
{A.\ Klemm, W.\ Lerche and P.\ Mayr,
 \nihil{K3 Fibrations and Heterotic-Type II String Duality,}
 Phys.\  Lett.\ {\bf B357} (1995) 313-322,
 \eprt{hep-th/9506112}}.}

\lref\HM{
{J.\ Harvey and G.\ Moore,
 \nihil{Algebras, BPS States, and Strings,}
 Nucl.\  Phys.\ {\bf B463} (1996) 315-368,
 \eprt{hep-th/9510182}.}}

\lref\gm{
{G.\ Moore, \nihil{String duality,
automorphic forms, and generalized
Kac--Moody algebras}, \eprt{hep-th/9710198};\br}
{M.\ Marino and G.\ Moore,
 \nihil{Counting higher genus curves in a Calabi-Yau manifold,}
 \eprt{hep-th/9808131}.}}

\lref \DKLII{L. Dixon, V. Kaplunovsky and J. Louis,
 \nihil{Moduli dependence of string loop corrections to gauge coupling
constants,} Nucl.\ Phys.\ {\bf B355} (1991) 649-688.}

\lref\fs{K.\ F\"orger and S.\ Stieberger,  {\nihil{String amplitudes
and $N=2$, $d=4$ prepotential in heterotic
$K3\times T^2$ compactifications,}
Nucl.\ Phys.\ {\bf B514} (1998) 135,  \eprt{hep-th/9709004}.}}

\lref\BK{C.\ Bachas and E.\ Kiritsis,
 \nihil{$F^4$ terms in N=4 string vacua,}
 Nucl.\  Phys.\  Proc.\  Suppl.\ {\bf 55B} (1997) 194,
 \eprt{hep-th/9611205}.}

\lref\borch{R.E.\ Borcherds, \nihil{Automorphic forms and
$O_{s+2,2}(\IR)$ and infinite products},
{Invent.\ Math.} {\bf 120} (1995) 161; \nihil{Automorphic forms with
singularities on Grassmannians,}
\eprt{alg-geom/9609022}. }

\lref\LY{
{B.\ Lian and S.\ Yau,
 \nihil{Arithmetic properties of mirror map and quantum coupling,}
 Commun.\  Math.\  Phys.\ {\bf 176} (1996) 163-192,
 \eprt{hep-th/9411234};} \br
{
 \nihil{Mirror maps, modular relations and hypergeometric series 1,}
 \eprt{hep-th/9507151};}
{
 \nihil{Mirror maps, modular relations and hypergeometric series. 2,}
 \eprt{hep-th/9507153}.}
}

\lref\elias{For a comprehensive review, see: E.\ Kiritsis,
 \nihil{Introduction to nonperturbative string theory,}
 \eprt{hep-th/9708130}.}

\lref\WL{W. Lerche, {
 \nihil{Elliptic index and superstring effective actions,}
 Nucl.\  Phys.\ {\bf B308} (1988) 102.}}

\lref\ellg {A.\ Schellekens and N.\ Warner,
{\nihil{Anomalies, characters and strings,}
 Nucl.\  Phys.\ {\bf B287} (1987) 317;}\br
{E.\ Witten,
 \nihil{Elliptic genera and quantum field theory,}
 Commun.\  Math.\  Phys.\ {\bf 109} (1987) 525;}\br
W. Lerche, B.E.W. Nilsson, A.N. Schellekens and N.P. Warner,
\np {\bf 299} (1988) 91.}

\lref\mirror{
See e.g.,
\nihil{Essays and mirror manifolds}, (S.\ Yau, ed.),
International Press 1992;
\nihil{Mirror symmetry II}, (B.\ Greene et al, eds.),
International Press 1997.
}

\lref\LS{W.\ Lerche and S.\ Stieberger,
 \nihil{Prepotential, mirror map and F-theory on K3,}
 \eprt{hep-th/9804176}.}

\lref\YZ{S.\ Yau and E.\ Zaslow,
 \nihil{BPS states, string duality, and nodal curves on K3,}
 Nucl.~ Phys.~{\bf B471} (1996) 503-512,
 \eprt{hep-th/9512121}.}

\lref\BL{
L.\ Bryan and N.\ Leung,
 \nihil{The enumerative geometry of K3 surfaces and modular forms,}
 \eprt{alg-geom/9711031}.}

\lref\BS{L.\ Baulieu and S.\ Shatashvili,
 \nihil{Duality from topological symmetry,}
 \eprt{hep-th/9811198}.}

\lref\Lee{M.\ Lee,
 \nihil{Picard-Fuchs equations for elliptic modular varieties,}
 Appl.\ Math.\ Letters 4, no.5 (1991) 91-95.
}

\lref\CD{C.\ Doran,
 \nihil{Picard-Fuchs Uniformization:
  Modularity of the Mirror Map and Mirror-Moonshine},
 \eprt{math.AG/9812162}.}

\lref\KLMVW{A.\ Klemm, W.\ Lerche, P.\ Mayr,
 C.\ Vafa and N.\ Warner,
 \nihil{Self-dual strings and N=2 supersymmetric field theory,}
 Nucl.\  Phys.\ {\bf B477} (1996) 746-766,
 \eprt{hep-th/9604034}.}

\lref\HKTY{S.\ Hosono, A.\ Klemm, S.\ Theisen and S.\ T.\ Yau,
 \nihil{Mirror symmetry, mirror map and applications to Calabi-Yau
hypersurfaces,}
 Commun.~ Math.~ Phys.~{\bf 167} (1995) 301-350,
 \eprt{hep-th/9308122}.}

\lref\specialG{
{A.\ Strominger,
 \nihil{Special geometry,}
 Commun.~ Math.~ Phys.~{\bf 133} (1990) 163-180};\br
S.\ Ferrara and A.\ Strominger, \nihil{N=2 spacetime
supersymmtry and Calabi-Yau moduli space,} in: Proceedings of
College Station Workshop (1989) 245.}

\lref\BD{A.\ Beauville and R.\ Donagi,
\nihil{La vari\'et\'e des droits d'une hypersuface cubique de
dimension 4,} C.\ R.\ Acad.\ Sci.\ Paris 301 (1985) 703-706.
}

\lref\LSW{W.\ Lerche, S.\ Stieberger and N.\ Warner,
 \nihil{Quartic gauge couplings from K3 geometry,}
 \eprt{hep-th/9811228}.}

\lref\fsII{K.\ Foerger, and S.\ Stieberger,
 \nihil{Higher derivative couplings
  and heterotic type I duality in eight-dimensions,}
 \eprt{hep-th/9901020}.}

\lref\WitM{E.\ Witten,
\nihil{Solutions of four-dimensional field theories via M-theory,}
 Nucl.\  Phys.\ {\bf B500} (1997) 3-42,
 \eprt{hep-th/9703166}.}

\lref\anom{
{A.\ N.\ Schellekens, N.\ P.\ Warner,
 \nihil{Anomalies and modular invariance in string theory,}
 Phys.~ Lett.~{\bf 177B} (1986) 317;\br}
{A.\ N.\ Schellekens, N.\ P.\ Warner,
 \nihil{Anomalies, characters and strings,}
 Nucl.~ Phys.~{\bf B287} (1987) 317.}
}

\lref\BSV{
M.\ Bershadsky, V.\ Sadov and C.\ Vafa,
\nihil{D-Branes and Topological Field Theories,}
\np {\bf 463} (1996) 420-434,
\eprt{hep-th/9511222}.}

%%%%%%%%%%%%%%%%%%%%%%%%%%%% %%%%%%%%%%%%%%%%%%%%%%%%%%%%%

%\draft

\def\pubnum{
\hbox{CERN-TH/99-17}
\hbox{hep-th/9901162}}
\def\pdate{}
\titlepage
\vskip2.cm
\title
{{\titlefont  Prepotentials from Symmetric Products}}
\vskip -.7cm
\autskip
\author{\wl,
S.~Stieberger and N.P.~Warner\footnote{*}{On leave
from Physics Department,
U.S.C., University Park, Los Angeles, CA 90089-0484}}
%%\andauthor{  }
\vskip0.2truecm
\CERN
\vskip0.2truecm
%\vskip-.8truecm

\abstract{
We investigate the prepotential that describes certain $F^4$ couplings
in eight dimensional string compactifications, and show how they can be
computed from the solutions of inhomogenous differential equations.
These appear to have the form of the Picard-Fuchs equations of a
fibration of Sym$^2(K3)$ over $\IP^1$. Our findings give support to the
conjecture that the relevant geometry which underlies these couplings
is given by a five-fold.}

\vfil
\vskip 1.cm
\ni {CERN-TH/99-17}\hfill\break
\ni January 1999
\endpage
\baselineskip=14pt plus 2pt minus 1pt

%\sequentialequations

%%%%%%%%%%%%%%%%%%%%%%%%%%%%%%%%%%%%%%%%%%%%%
\chapter{Introduction}
%%%%%%%%%%%%%%%%%%%%%%%%%%%%%%%%%%%%%%%%%%%%%

In string theories with extended supersymmetry, BPS-saturated
amplitudes \multref\HM{\BK}\elias\ play an important r\^ole for
non-trivial tests of various kinds of dualities. They tend to be
characterized by holomorphic quantities (\eg\ prepotentials), and this
is why one often can use geometrical methods to compute them exactly.
Typically, the counting of BPS states that contribute to a given
amplitude can be mapped to the computation of the Euler characteristic
of a space of geometric moduli. For prepotentials this generically
reduces to the counting of curves in some complex manifold $X$, and
this manifold may, or may not have a concrete physical meaning in some
appropriate dual formulation of the theory. In practice, this counting
is often done via mirror symmetry \mirror, which boils down to
computing the

Some of the most canonical BPS-saturated amplitudes involve an even
number, $n$, of external gauge bosons in theories with $4n$
supercharges in $2n$ dimensions. These amplitudes arise in heterotic
string compactifications on $Y\times T^2$, where $Y$ is some
$(4-n)$-fold. In the following, we will focus only on the subsector of
the theory that depends on the familiar torus moduli $T$ and $U$
(neglecting any Wilson lines), and consider couplings of the form
$\Delta_{{F_T}^{n-m}{F_U}^m}(T,U) F_T\wedge ...F_T  \wedge F_U
\wedge ...F_U$, which are saturated by 1/2-BPS states. In the heterotic
string formulation, the perturbative piece is given by a one-loop
amplitude that involves \WL\ the heterotic elliptic genus \ellg\
$\ellgen_{-n}$ in $2n+2$ dimensions, e.g.,
$$
\Delta_{{F_T}^n}=
\int {d^2\tau \o \tau_2}\ \sum_{(p_L,p_R)}\ov p_R^n\ q^{\h|p_L|^2}\ov
q^{\h|p_R|^2}\ \ellgen_{-n}(\ov q)\ .
\eqn\done
$$
Here, $p_L={1\o \sqrt{2T_2 U_2}}(m_1+m_2\bar U+n_1\bar T+n_2\bar T\bar
U)$ and $p_R={1\o \sqrt{2T_2 U_2}}(m_1+m_2\bar U+n_1 T+n_2 T\bar U)$
are the usual Narain momenta of the compactification torus $T^2$.

By explicitly performing the modular integral in \done\ for general
$n$, we find (by extensive calculations generalizing methods developed
in \nrf{\DKLII{\fs \LS }\fsII} \refs{\DKLII,\HM,\fs{--}\fsII}) that
these couplings satisfy non-trivial integrability conditions.\foot{An
explicit demonstration of this for $n=6$ is given in Appendix A.} These
imply that the couplings $\Delta_{{F_T}^{n-m}{F_U}^m}(T,U)$ can be
written as $n$-fold (covariant) derivatives with respect to $T,U$ of
the following holomorphic prepotentials:
$$
\eqalign{
\Fn n(T,U)\ =\ &-(-1)^{n/2} {i c\lb n(0)
\zeta(n+1)\o 2^{n+2}\pi^{n+1}}-
{U^{n+1} \o (n+1)!}+\cQ(T,U)
\cr & -
(-1)^{n/2} {i \o (2\pi)^{n+1}}\sum_{(k,l)>0} c\lb n(k \,
l)\Li_{n+1}\lf[{q_T}^k{q_U}^l\ri].
}\eqn\prep
$$
Here, $\cQ(T,U)$ is some undetermined $n$-th order polynomial in $T,U$
and $TU$ (with real coefficients), $q_T\equiv e^{2\pi i T}$, $q_U\equiv
e^{2\pi i U}$, and  $\Li_a(z)=\sum_{p>0} {z^p \o p^a}$  is the $a$-th
polylogarithm.  The  sum runs over the positive roots $k>0,\ l\in \ZZ\
\ \wedge\ \ k=0,\ l>0$, and the coefficients, $c\lb n$, are simply the
expansion coefficients of the corresponding elliptic genus,
$\ellgen_{-n}(q)=:\sum_{k\geq -1}c\lb{n}(k)q^k$, which is a modular
form of weight $-n$.

Of course, for $n=2$\ (\ie, $N=2$ supersymmetry in four dimensions) the
situation is well understood; the prepotential is nothing other
than the effective lagrangian of special geometry \specialG. A dual
formulation is given by Type II A/B strings compactified on the
familiar $K3$-fibered Calabi-Yau threefold
$X_{24}(1,1,2,8,12)^{-480}_{3}$, and its mirror. The mirror symmetry
allows to exactly compute the full non-perturbative prepotential
$\FSTU\lb2(S,T,U)$, which also involves the dilaton modulus,
$S$. The one-loop prepotential $\Fn2(T,U)$ in \prep, with
$$
\ellgen_{-2}(q)\ \equiv {E_4E_6\over \eta^{24}}(q)\ ,
\eqn\gentwo
$$
is then reproduced \KLM\ in the weak coupling limit, $S\to\infty$,
where the non-perturbative corrections disappear.

On the other hand, the situation is much less well understood\foot{Not
the least because an appropriate generalization of special geometry, in
which $\cF^{(4)}(T,U)$ would figure as a superspace lagrangian, is not
known. However, see \BS\ for some recent progress in eight dimensional
lagrangians.} for $n=4$, which corresponds to $N=1$ supersymmetry ($16$
supercharges) in eight dimensions, and where
$$
\ellgen_{-4}(q)\ \equiv {{E_4}^2\over \eta^{24}}(q)\ .
\eqn\genfour
$$
An interesting issue is to find a geometrical computation that would
lead to the prepotential $\cF^{(4)}(T,U)$, in an analogous manner to
the more familiar computation that leads to $\cF^{(2)}(T,U)$.

Since the dual formulation of the eight-dimensional heterotic
compactification on $T^2$ is given by $F$-theory \Fth\ compactified on
$K3$, one would expect that $\cF^{(4)}(T,U)$  should be computable in
terms of the geometrical data of $K3$.  The main puzzle is that the
prepotential $\cF^{(4)}(T,U)$ does not seem to be in any obvious way
related to $K3$, but rather looks like a prepotential that
would canonically come from a five-fold. This is essentially because
its fifth derivatives have exactly the structure as ``world-sheet
instanton corrected Yukawa couplings'', \ie\ ${\del_T}^m
{\del_U}^{5-m}\cF^{(4)}(T,U)={\rm const}+\sum_{k,l}c \lb4 (k\,l) k^m
l^{5-m}{ {q_T}^k{q_U}^l\over 1- {q_T}^k{q_U}^l}$.

Some preliminary investigations in this direction have been presented
in \doubref\LS\LSW, and in particular in \LSW\ evidence was found
that the relevant five-fold should be given by the symmetric square,
Sym$^2(K3)$, fibered over $\IP^1$ (where the size of $\IP^1$ is
eventually taken to be infinite). This structure was uncovered by
investigating certain other couplings (involving four external
non-abelian gauge fields), for which no prepotential exists. It is the
purpose of the present paper to extend this analysis to the couplings
$\Delta_{{F_T}^m{F_U}^{4-m}}$ and their prepotential $\cF^{(4)}(T,U)$,
and gather further evidence that the relevant underlying quantum
geometry is given by such a five-fold.

Here we will not, however, try to answer the question as to what the
physical interpretation of this five-fold might be, if there is any at
all. The situation is, in this respect, somewhat similar to $N=2$ SYM
theory in four dimensions, where the Riemann surfaces underlying the
effective lagrangian were found in \SW, and at the time the geometry
appeared to be merely a convenient mathematical tool for encoding
appropriate data. It was only later that the geometry was given a much
deeper physical interpretation.\foot{For example, as part of
world-volumina of type IIA \KLMVW\ or $M$-theory \WitM\ fivebranes.}
In the same spirit, one may speculate that the five-folds that seem to
emerge here may ultimately have an interpretation in terms of a yet
unknown dual formulation of the theory, or, perhaps more likely, in
terms of sigma-models describing the relevant 7-brane interactions \LS\
that lead to the requisite $F^4$ terms in the effective action. Indeed,
sigma models on symmetric products of $K3$ do naturally appear in
$D$-brane physics \BSV, so that there is hope that we may eventually
learn something substantially new about how to do exact
non-perturbative computations.

In the next section, we will review how the perturbative prepotential
$\cF^{(2)}(T,U)$ arises geometrically; in particular, we will derive
the inhomogenous Picard-Fuchs equations that capture the relevant
information of the $K3$ fibration in the large base space limit. The
motivation is, of course, to subsequently apply the {\it reverse} of
this procedure to the eight-dimensional situation, where we want to
start from the known perturbative prepotential $\cF^{(4)}(T,U)$, to
arrive at a large base space limit of some fibration. This will be done
in section 3, where we will find that the periods of the fiber are
given by the squares of the ordinary $K3$ periods, \ie\ by
$(1,T,U,TU,T^2,U^2,T^2U^2)$. These are precisely the periods of the
hyperk\"ahler symmetric square of $K3$, which we denote by
Sym$^2(K3)$.  In the appendix, we formally extend this reasoning  to
$n=6$ external gauge bosons, and relate $\cF^{(6)}(T,U)$ to cubic
powers of the $K3$ periods. More generally, we conclude that the
prepotentials $\cF^{(n)}(T,U)$ can be formally related to
$(n+1)$-folds, given by $\IP^1$ fibrations of symmetric products,
Sym$^{n/2}(K3)$. Finally, we will present some comments on curve
counting in $K3$.

%%%%%%%%%%%%%%%%%%%%%%%%%%%%%%%%%%%%%%%%%%%%%%%%%%%%%%%%%%%
\chapter{The Prepotential $\cF^{(2)}$ in the Large Base-Space Limit}
%%%%%%%%%%%%%%%%%%%%%%%%%%%%%%%%%%%%%%%%%%%%%%%%%%%%%%%%%%%

\ni The defining
polynomial of the Calabi-Yau manifold
$X_{24}(1,1,2,8,12)^{-480}_{3}$ is given by
$$
p = x_1^2 + x_2^3 + x_3^{12} + x_4^{24} + x_5^{24} -
12 \psi_0 x_1x_2x_3x_4x_5-2\psi_1 (x_3x_4x_5)^6-
\psi_2(x_4x_5)^{12}\, .
\eqn\npolI
$$
As described in \KLM, this Calabi-Yau manifold may be thought of as a
fibration of a  $K3$ family of type $X_{12}(1,1,4,6)$ over  the $\IP^1$
base defined by the coordinates $x_1,x_2$. Moreover  this $K3$ is
itself an elliptic fibration over $\IP^1$ with  generic fiber
$X_6(1,2,3)$.

The variables that are appropriate for describing the complex
structure near the point of maximal unipotent monodromy in the large
complex structure limit are: $x=-{2\psi_1\over 1728^2\psi_0^6}$,
$y={1\over\psi_2^2}$, $z =-{\psi_2\over 4\psi_1^2}$.  In these
variables the Picard-Fuchs (PF) system, which determines the three-fold
periods, becomes \HKTY
$$
\eqalign{
{\cal D}_1\cy &~=~\tx\,(\tx-2\,\tz)-12\, x\,(6\,\tx+5)\,(6\,\tx+1)
\ , \cr
{\cal D}_2\cy &~=~\tz \,(\tz-2\ty)- z (2\, \tz-\tx +1)\,(2\, \tz-\tx )
\ ,\cr
{\cal D}_3\cy &~=~\ty^2- y\, (2\,\ty-\tz+1)\,(2\,\ty-\tz) \ ,}
\eqn\PFthree
$$
where $\theta_x \equiv x {d \over d x}$ etc.
For $y\rightarrow 0$ this system degenerates to the two moduli system
of the $K3$ fiber:
$$
\eqalign{
{\cal D}_1\kt ~=~ &\tx^2\, -12\, x\,(6\,\tx+5)\,(6\,\tx+1)\ , \cr
{\cal D}_2\kt ~=~ &\tz^2\, - z (2\, \tz-\tx +1)\,(2\, \tz-\tx )
 \ . }\eqn\PFtwo
$$

Denoting the flat coordinates by $S,T$ and $U$, in the usual
manner, the prepotential of this Calabi-Yau manifold can be
written in the form
$$
\FSTU\lb2(S,T,U)  ~=~ S T U ~+~ \Fn2(T,U) ~+~
\sum_{n=1}^\infty g_n(T,U) ~ {q_S}^n    \ ,  \eqn\genericF
$$
where $q_S = e^{-4\pi S}$, and $y \sim q_S$ as $S \to \infty$.  In this
expression, the first term is the classical part, and the second term,
$\Fn2(T,U)$, may be thought of as the perturbative one-loop part of the
prepotential that comes from the $K3$ fiber. The last sum is over
world-sheet instantons that wrap  the base, which gives the
non-perturbative corrections from the heterotic string point of view.
Our aim is  to extract the function $\Fn2(T,U)$, and compare\foot{Of
course, this has been already done before in \KLM; our purpose here
is to formulate the problem in a way that allows an easy
generalization to eight dimensions.} it with the heterotic one-loop
prepotential given in \prep. To do this we must carefully take the
limit  $S \to \infty$  in the PF system, keeping track all the
divergent and finite parts.

Let $\pi_0$ and $\varpi_0$ be the fundamental periods  of the
Calabi-Yau and the $K3$, respectively.  They are the  unique solutions
of \PFthree\ and \PFtwo\ with finite limits  at $x=0$ and $z=0$.  Then
the following  represents the asymptotics (as $S \to \infty$) of
the Calabi-Yau three-fold periods:
%$$
%\eqalign{ S \pi_0 ~\sim~ & ~ (\log(y)  ~+~ \mu_0(T,U))\varpi_0
%\ , \qquad  T \pi_0 ~\sim~ T \varpi_0 \ , \cr
%U \pi_0 ~\sim~& ~ U \varpi_0  \ , \qquad {\cal F}_S \pi_0 ~\sim~
% T U \varpi_0 \ , \cr
%{\cal F}_T \pi_0 ~\sim~ & U(\log(y)  ~+~ \mu_0(T,U) ~+~ f_T (T,U))
%\varpi_0 \ , \cr {\cal F}_U \pi_0 ~\sim~ & T(\log(y)  ~+~
%\mu_0(T,U)   ~+~  f_U (T,U)) \varpi_0  \ , \cr
%{\cal F}_0 \pi_0 ~\sim~ & T U(\log(y) ~+~ \mu_0(T,U)  ~+~
%f_0 (T,U) )\varpi_0
%\ . } \eqn\asymptotics
%$$
$$
\eqalign{
\pi_0 ~ &\sim ~  \varpi_0\cr
 T \pi_0 ~&\sim~ T \varpi_0  \cr
U \pi_0 ~&\sim~ U \varpi_0 \cr
\FSTU\lb2_S \pi_0 ~&\sim~
 T U \varpi_0 \ , \cr
S \pi_0 ~&\sim~  (\log(y)  ~+~ \mu_0(T,U))\varpi_0 \cr
\FSTU\lb2_T \pi_0 ~&\sim~ (U(\log(y)  ~+~ \mu_0(T,U))~+~ \Fn2_T (T,U))
\varpi_0  \cr
\FSTU\lb2_U \pi_0 ~&\sim~  (T(\log(y)  ~+~
\mu_0(T,U))  ~+~  \Fn2_U (T,U)) \varpi_0   \cr
\FSTU\lb2_0 \pi_0 ~&\sim~  (T U(\log(y) ~+~ \mu_0(T,U))  ~+~
\Fn2_0 (T,U))\varpi_0
\ . } \eqn\asymptotics
$$
We see that in this limit, the first four CY periods turn directly into
the periods of the $K3$ fiber, which are the solutions of \PFtwo. On
the other hand, the non-trivial function that we seek, $\Fn2(T,U)$, is
encoded in the remaining half of  the periods. These are governed by an
{\it inhomogenous} Picard-Fuchs system \LSW, whose homogenous part is
exactly the system \PFtwo\ of the $K3$ fiber, and whose source part
stems from $\ty$ in ${\cal D}_2\cy$ hitting log$(y)$ (which survives
the $y\to0$ limit).  More precisely,  it follows from \PFthree\ and
\asymptotics\ that if  $\mu_{jk}$ are the solutions to
$$
{\cal D}_1\kt (\mu_{jk} \varpi_0) ~=~ 0
 \ ,   \qquad {\cal D}_2\kt  (\mu_{jk} \varpi_0) ~=~
T^j~U^k \, (\tz \, \varpi_0) \ ,
\eqn\PFsource
$$
then we have
%$$
%\eqalign{
%\mu_0 ~=~& \mu_{00}\ , \qquad U \mu_{00} ~=~ \mu_{01} + \Fn2_T  \ ,\cr
%\Fn2_U ~=~&    \mu_{10} - T \mu_{00} \ , \qquad
%\Fn2_0 ~=~  \mu_{11} - TU \mu_{00} \ ,
%}\eqn\relns
%$$
$$
\eqalign{
\mu_0 ~=~& \mu_{00}\ , \qquad \mu_{01}~=~ \Fn2_T +  U \mu_{00}\ ,\cr
\mu_{10}~=~&\Fn2_U +  T \mu_{00}\ , \qquad
\mu_{11}~=~ \Fn2_0 +  TU \mu_{00}\ ,
}\eqn\relns
$$
and in particular, from homogeneity:
$$
\Fn2(T,U) ~=~\mu_{11} - T \mu_{01}
- U \mu_{10} + TU \mu_{00}  \ ,
\eqn\FfromPF
$$
which reflects the familiar relation $\cF=\shalf X^A\cF_A$
of special geometry.

To explicitly see that \FfromPF\ indeed coincides with the heterotic
one-loop expression \prep, we first need to simplify the PF system
\PFthree. To accomplish this, we make a change of variables to $w_1,
w_2$, where:
$$
\eqalign{x~=~& {1 \over 864}~\Big[~1~-~ \sqrt{(1- w_1)~
(1- w_2)} \Big] \ , \cr
z~=~& {w_1 w_2 \over 4}~ (w_1 + w_2 - w_1 w_2)^{-2}~\Big[~1~+~
\sqrt{(1- w_1)~  (1- w_2)} \Big]^2 \ .}
\eqn\chofvars
$$
{}From the explicit expressions given in \KLM\ it follows that simply
$$
w_1={1728\over j(T)}\ , \ \qquad\ w_2={1728\over j(U)}\ .
\eqn\simplemirrormap
$$
This effectively separates variables in the PF equations, and  one
finds
%$$
%\eqalign{
%{\cal D}_1\kt ~=~  & {1728~x \over w_1 - w_2} ~\Big[w_1~
%\cL\ow_{w_1}~-~ w_2~\cL\ow_{w_1} \Big]\ , \qquad
%{\cal D}_2\kt ~=~  - {w_1  w_2 \over w_1 - w_2} ~\Big[
%\cL\ow_{w_1}~-~ \cL\ow_{w_2} \Big] \cr &
%\tz ~=~  - {w_1  w_2 \over w_1 - w_2}~\Big[~(1-w_1) {d \over
%d w_1} ~-~ (1-w_2) {d \over d w_2}~ \Big] \ , }\eqn\PFtwosep
%$$
$$
\eqalign{
{\cal D}_1\kt ~&=~   {1728~x \over w_1 - w_2} ~\Big[w_1~
\cL\ow_{w_1}~-~ w_2~\cL\ow_{w_1} \Big]\cr
{\cal D}_2\kt ~&=~  - {w_1  w_2 \over w_1 - w_2} ~\Big[
\cL\ow_{w_1}~-~ \cL\ow_{w_2} \Big] \cr
\tz ~&=~  - {w_1  w_2 \over w_1 - w_2}~\Big[~(1-w_1) {d \over
d w_1} ~-~ (1-w_2) {d \over d w_2}~ \Big] \ , }\eqn\PFtwosep
$$
where $\cL\ow_w$ is the second order hypergeometric operator
$$
{\cal L}_w\ow  ~\equiv~  {1 \over w}~
\Big[ \theta_w^2 ~-~ w~(\theta_w + {5 \over 12})(\theta_w +
{1 \over 12}) \Big]\ .
\eqn\Ltwo
$$
The fundamental period $\varpi_0$ of the $K3$ must therefore
satisfy $\cL\ow_{w_1} \varpi_0 = \cL\ow_{w_2} \varpi_0 = 0$,
and hence it must have the form  $\varpi_0 = \omega_0 \tilde
\omega_0$, where $\omega_0$ is given by the fundamental series solution
of \Ltwo:
$$
\omega_0 (w) ~=~ {}_2F_1\Big({1 \over 12},
{5 \over 12};1,w\Big)~=~ (E_4)^{1/4} \ .
\eqn\fundsersol
$$
with $w=w_1$, and $\tilde \omega_0$ is the same function but with
$w=w_2$.
Using \PFtwosep\ the equations \PFsource\ can be rewritten as
$$
\cL\ow_{w_a} (\mu_{jk}~\varpi_0) ~=~ - {1 \over w_a}~
{2 w_1  w_2 \over w_1 - w_2}~\Big[~(1-w_1) {d \over
d w_1} ~-~ (1-w_2) {d \over d w_2}~ \Big] (T^j U^k \varpi_0) \ .
\eqn\PFsourcesep
$$
{}From this and \relns\ it follows, for example, that:
$$
w_1 \cL\ow_{w_1} (\Fn2_T~\varpi_0) ~=~ {w_1 w_2 \over w_1 - w_2}~
(1-w_2)~{d U \over d w_2}~ \varpi_0 \ .
$$
Using \simplemirrormap\ and the identity\foot{This identity
is straightforward and is simply the result of a the
change of variables~\simplemirrormap.} \LSW
$$
w~{\cal L}\ow_w~(f(w)\omega_0(w)) ~=~
{1 \over E_4(T)}~(\theta_{q_T}^2~f(w(T))) ~\omega_0\
\eqn\crucialident
$$
(for any function $f(w)$), we finally see that:
$$
(\theta_{q_T}^2 ~\Fn2_T ) ~=~ -E_4(T)~{w_1 w_2 \over w_1 - w_2}~
(1-w_2)~ {d U \over d w_2}~=~ { E_4(T)~E_4(U) E_6(U)\over
[j(T) - j(U)] \eta^{24}(U)} \ .
\eqn\FTTT
$$
This coincides exactly with the known
\doubref\oneloop\KLM\ expression for $\Fn2_{TTT}(T,U)$.

Summarizing, we have shown how the perturbative component of the
quantum prepotential can be obtained directly from the $K3$
Picard-Fuchs equations with properly chosen sources, and these sources
are simply derivatives of the $K3$ periods.

We now briefly indicate how to reverse this process, and in the next
section we will use this method to construct differential equations
whose solutions lead to the other $\Fn n(T,U)$.

It turns out that the strongest single constraint on the form of the
differential operators comes from the explicit form of the dilaton,
which is essentially the difference (at large $S$) between the solution
$\mu_{00}$ and  the manifestly modular invariant quantity log$(y)$.
Since the dilaton is non-singular at $T=U$, this solution must have the
form
$$
\mu_{00} ~=~ 2 \pi i~\Fn2_{TU} - \log(j(T) - j(U)) \ .
\eqn\invdil
$$
The general idea is to first obtain a differential equation for
$\Fn2_{TU}$, by inserting it into the identity \crucialident\ with
$w=w_1(T)$. The right-hand side of this equation, which represents the
source part, is then given by $(1/E_4(T))(\del_U\Fn2_{TTT})\omega_0$,
which can be evaluated by using the known expression \FTTT\ for
$\Fn2_{TTT}(T,U)$. After subtracting the logarithmic singularity, this
leads precisely to the source term on the RHS of the Picard-Fuchs
system \PFsourcesep.

%%%%%%%%%%%%%%%%%%%%%%%%%%%%%%%%%%%%%%%%%
\chapter{Generalizations}
%%%%%%%%%%%%%%%%%%%%%%%%%%%%%%%%%%%%%%%%%

Assuming that the Picard-Fuchs equations we seek for $n=4$ generalize
the structure we found above, we will try to construct differential
equations for $\Fn n_{T^{n/2}U^{n/2}}(T,U)$ by applying  the simple
procedure outlined earlier. However, before doing that, we will first
discuss some general features of the homogenous PF equations for
arbitrary~$n$.

%%%%%%%%%%%%%%%%%%%%%%%%%%%%%%%%%%%%%%%%%
\subsec{Symmetric Powers of Picard-Fuchs Operators}
%%%%%%%%%%%%%%%%%%%%%%%%%%%%%%%%%%%%%%%%%

Crucial to our arguments will be the following sequence of
differential operators:
$$
\eqalign{
{\cL}_w\ot1  ~\equiv~ & {1 \over w}~
\Big[ \theta_w^2 ~-~ w~(\theta_w + {5 \over 12})(\theta_w +
{1 \over 12}) \Big]  ~\equiv~\cL\ow_w \ , \cr
{\cL}_w\ot2  ~\equiv~ & {1 \over w}~
\Big[ \theta_w^3 ~-~ w~(\theta_w + {5 \over 6})(\theta_w +
{1 \over 2})(\theta_w +  {1 \over 6}) \Big] \ , \cr
{\cL}_w\ot3  ~\equiv~ & {1 \over w}~ \Big[ \theta_w^4 ~-~
2 w ~(\theta_w + {1 \over 4})(\theta_w^3  + {5 \over 4} \theta_w^2 +
{31 \over 36}  \theta_w +  {5 \over 24})
\cr &  \qquad \quad ~+~
w^2 ~(\theta_w + {5 \over 4})(\theta_w + {11 \over 12})
(\theta_w +  {7 \over 12}) (\theta_w +  {1 \over 4})\Big]
\ , \quad {\rm etc.,}}
\eqn\threeDOs
$$
where ${\cL}_w\ot1\equiv{\cL}_w\ow$ is identical to the hypergeometric
operator in \Ltwo. The $(m+1)^{\rm th}$ order operator ${\cL }_w\ot m$
is what is called the ``$m^{\rm th}$ symmetric power'' of the basic
operator ${\cL}_w$, the reason being that its solution space is the
$m^{\rm th}$ symmetric product of the solution space of ${\cL}_w\ow$.
The notion of symmetric powers of differential operators has been
discussed in the mathematical literature, \eg\  in \LY\ and in
\doubref\Lee\CD, where also a systematic procedure for computing them
has been described.

More explicitly, while the fundamental solutions to  ${\cal
L}_w\ow\omega_i(w) =0$  are given by the periods
$$
\omega_0 (w) ~=~ {}_2F_1\Big({1 \over 12},{5 \over 12};1,w\Big)~=~
(E_4)^{1/4}\ , \quad \omega_1 (w) ~=~ T~\omega_0 ~=~ T~(E_4)^{1/4}\ ,
\eqn\basicperiods
$$
the solutions of ${\cL}_w\ot m$ are given by
$$
\eqalign{
\omega_j\ot 2(w) ~=~ & \omega_{j-i}~\omega_{i} ~=~ T^j~(E_4)^{1/2}\ ,
\qquad j=0,1,2 \cr
\omega_k\ot 3(w) ~=~ & \omega_{k-j}~\omega_{j-i}~\omega_i ~=~
T^k~(E_4)^{3/4}\ ,  \qquad i,j,k=0,1,2,3 \ , }
\eqn\polyrels
$$
and so on. Moreover, we find that these operators satisfy certain
identities when filtered through the mirror  map, $w=1728/j(T)$: for
any function $f(z)$ one has
$$
\eqalign{w~{\cL}\ot1~(f(w)\omega_0(w)) ~=~
&{1 \over E_4(T)}~({\theta_{q_T}}^2~f(w(T))) ~\omega_0\ , \cr
w~{\cL}\ot2~(f(w){\omega_0}(w)^2) ~=~
&{1 \over E_6(T)}~({\theta_{q_T}}^3~f(w(T))) ~{\omega_0}^2\ , \cr
w~{\cL}\ot3~(f(w){\omega_0}(w)^3) ~=~
&{1 \over E_4^2 (T)}~({\theta_{q_T}}^4~f(w(T))) ~{\omega_0}^3
\ ,\quad {\it etc.}}
\eqn\diffidents
$$
These identities will prove important momentarily.

%%%%%%%%%%%%%%%%%%%%%%%%%%%%%%%%%%%%%%%%%
\section{Determination of the source terms}
%%%%%%%%%%%%%%%%%%%%%%%%%%%%%%%%%%%%%%%%%

Note that the prepotentials \prep\ have the property
that $\partial_T^{n+1}\Fn n(T,U)$ is a good  modular function of
weights $(n+2,-n)$ in $(T,U)$, and must have a simple pole at $T=U$
(which reflects gauge symmetry enhancement to $SU(2)$). From this one
can deduce the functional form. For example, one has:
$$
\eqalign{
\partial_T^3 ~\Fn2(T,U) ~=~& {E_4(T) ~ E_4(U) E_6(U)
\over  [J(T) ~-~ J(U)]~\eta^{24}(U)} \ , \cr
\partial_T^5 ~\Fn4(T,U) ~=~
& {E_6(T) ~ E_4^2(U) \over [J(T) ~-~ J(U)]~\eta^{24}(U)} \ , \cr
 \partial_T^7 ~\Fn6(T,U) ~=~&
{E_4^2(T) ~ E_6(U) \over[J(T) ~-~ J(U)]~\eta^{24}(U)} \ ,
\quad{\it etc.} }
\eqn\derivpreps
$$
Suppose we set $w=w_1$ in \diffidents, and take the $f$ to be $\del_T^m
\del_U^m f\lb{2m}(T,U)$. The right-hand side of the $m^{\rm th}$
equation in \diffidents\ can then be rewritten using $m$
$U$-derivatives of the $m^{\rm th}$ identity in \derivpreps. The
resulting right-hand side is completely modular of $T$, and almost
modular in $U$. Indeed, the right-hand side is $m^{\rm th}$ order in
$E_2(U)$. These factors of $E_2$ may be traded for derivatives of the
fundamental periods as follows:
One first notes that the fundamental periods of the
various PF systems can be written as $\varpi_0^m\equiv {\omega_0}^m
{\tilde\omega_0}^m$, where ${\omega_0}^m\equiv {E_4(T)}^{m/4}$ and
${\tilde\omega_0}^m\equiv {E_4(U)}^{m/4}$. Therefore, one can
express the $w_2$-derivatives of the periods in terms of $U$
derivatives to obtain:
$$
\twb {\tilde\omega_0}^m = {m {\tilde\omega_0}^m\over 4 E_4}\,
\twb (E_4) = {m {\tilde\omega_0}^m \over 12 E_4}~ w_2
\Big({d w_2 \over d U} \Big)^{-1} (E_2 E_4 - E_6) =
{m {\tilde\omega_0}^m \over 12}
 \bigg({E_2 E_4 \over E_6} - 1\bigg).
$$
More generally, $(\twb)^p~{\tilde\omega_0}^m$  may be written
in terms of a polynomial of degree $p$ in $E_2(U)$.  Conversely,
a polynomial of degree $p$ in $E_2(U)$ may be expressed
as a linear differential operator of order $p$ in $w_2$,
acting on $\varpi_0^m$.  In this way, one can use \derivpreps\
and  \diffidents\ to determine the right-hand sides of
$\cL\ot{(n/2)}_{w_a}\big[{\del_T}^{n/2}{\del_U}^{n/2} \Fn n
\,\varpi_0^{n/2}\big]$. The resulting  expressions have poles in
$(w_1-w_2)$ of orders  up to $(n/2+1)$.  To arrive at a PF system
similar to \PFtwosep\ one can tolerate at most  single poles in the
source terms (as in \PFsourcesep; this ensures that the ``dilaton''
period will be non-singular at $T=U$, \cf\ eq.~\invdil). The leading
pole can be cancelled by the addition of a  suitable multiple of
$\log(w_1 -w_2)$. The subleading poles  can then be cancelled by the
addition of multiples of ${\del_T}^{k/2}{\del_U}^{k/2} \Fn k$ for
$k<n$.

At the end of this iterative procedure, one arrives at
a pair of inhomogenous Picard-Fuchs equations of the general form,
$$
\cL\ot{(n/2)}_{w_a}\cdot\mu_{00}\lb n{\varpi_0}^{n/2}\ =\
\cM\lb{n/2}_a\cdot {\varpi_0}^{n/2}\ ,\qquad a=1,2,
\eqn\generalform
$$
which generalizes \PFsourcesep\ and whose source part involves some
$(n/2)^{\rm th}$-order operators $\cM\lb{n/2}_a$.  The homogenous,
``fiber'' part consists of two copies of the symmetric product of
$\cL_w$, whose solutions look, after dividing out the fundamental
period ${\varpi_0}^{n/2}$, like
$$
(1,T,U,TU,T^2,U^2,..., (TU)^{n/2})\ .
\eqn\homsols
$$
These are the periods of the $n/2$-fold symmetric product,
Sym$^{n/2}(K3)$.

%%%%%%%%%%%%%%%%%%%%%%%%%%%%%%%%%%%%%%%%%
\subsec{Explicit Results for $n=4$}
%%%%%%%%%%%%%%%%%%%%%%%%%%%%%%%%%%%%%%%%%

By following the steps described above, we find for
$n=4$ (which corresponds to the eight-dimensional
compactification) that
$$
\mu_{00}\lb 4 ~=~  {2 \pi i(\Fn4_{TTUU} ~+~ 3\Fn2_{TU}) ~-~
2 \log(w_1 -w_2) \ ,}
$$
satisfies the following inhomogenous PF equation:
$$
\cL\ot2_{w_1} \cdot\mu_{00}\lb 4{\varpi_0}^{2} ~=~
{6  w_2 \over (w_1 - w_2)}~ \bigg[\cL\ow_{w_1} ~+~
\cL\ow_{w_2} ~+~  w_1(1 -  w_2){ d^2
\over dw_1 dw_2}  ~-~ \coeff{5}{72} \bigg]\cdot {\varpi_0}^{2}  \ ,
\eqn\GPFsource
$$
along with the corresponding equation for $\cL\ot2_{w_2}(\mu_{00}\lb
4{\varpi_0}^{2})$ obtained by interchanging $w_1$ and $w_2$.
Since \GPFsource\ only involves a simple pole in
$w_1-w_2$, one can take sums or differences of the equations for
$\cL\ot2_{w_1}$ and $\cL\ot2_{w_2}$ so as to cancel the pole, and
obtain a form that more closely resembles the PF system of a manifold.
In particular, one can write:
$$
\eqalign{\big(w_1 \cL\ot2_{w_1}~+~ w_2 \cL\ot2_{w_2}\big)
\cdot\mu_{00}\lb 4{\varpi_0}^{2} ~=~
& 6 { d^2  \over dw_1 dw_2 }{\varpi_0}^{2} \ , \cr
\big((1-w_1)\cL\ot2_{w_1}+ (1- w_2)\cL\ot2_{w_2}\big)
\cdot \mu_{00}\lb 4{\varpi_0}^{2} ~=~
& 12~ \bigg[ \cL\ow_{w_1} ~+~
\cL\ow_{w_2}  - \coeff{5}{72} \bigg] \cdot {\varpi_0}^{2}
\ . } \eqn\otherform
$$

Having now obtained equations for the ``fundamental'' inhomogenous
solution  $\mu_{00}\lb4$, we can now investigate the full set of
solutions $\mu_{jk}\lb4$, for which ${\varpi_0}^{2}$ on the right-hand
side of \GPFsource\ or \otherform\ is replaced by
$T^j U^k{\varpi_0}^{2}$.
%$$
%\mu_{ij}\lb 4 ~\equiv~ T^j U^k~\rho_0\ .
%$$
One can then verify that the partial derivatives  of $\Fn 4$ are
related to  $\mu_{jk}\lb 4$ in a manner completely analogous to
\relns. Explicitly, abbreviating $\mu\equiv \mu\lb4$, one has:
$$
\eqalign{
& \mu_{01} - U \mu_{00}~=~ -6 \pi i~(\Fn2_{T} +\Fn4_{TTU})
 \cr &\mu_{10} - T \mu_{00}  ~=~ - 6 \pi i~(\Fn2_{U} +
\Fn4_{TUU}) \cr  & \mu_{02} - 2 U \mu_{01} + U^2 \mu_{00}~=~
-24 \pi i ~\Fn4_{TT}  \cr
&\mu_{20} - 2 T \mu_{10} + T^2 \mu_{00} ~=~ -24 \pi i ~\Fn4_{UU}
\cr  & \mu_{11} -  U \mu_{10} -  T \mu_{01}  + TU \mu_{00} ~=~
-6 \pi i ~(\Fn2_{} + 3 \Fn4_{TU}) \cr &
(\mu_{12} -  2 U \mu_{11} + U^2 \mu_{10}) - T (\mu_{02} -  2 U \mu_{01}
+
U^2 \mu_{00}) ~=~  - 72 \pi i ~\Fn4_{T}   \cr &
(\mu_{21} -  2 T \mu_{11} + T^2 \mu_{01}) - U (\mu_{20} -  2 T \mu_{10}
+
T^2 \mu_{00}) ~=~  - 72 \pi i ~\Fn4_{U}\ ,  \cr
} \eqn\Gtworelns
$$
and in particular:
$$
\eqalign{
216 \pi i ~\Fn4(T,U) \ &=\
(\mu_{22} -  2 U \mu_{21} + U^2 \mu_{20}) - 2 T (\mu_{12} -  2 U
\mu_{11}
+ U^2 \mu_{10}) + \cr &  \qquad \qquad
T^2 (\mu_{02} -  2 U \mu_{01} + U^2 \mu_{00})\ .
}\eqn\newspecialgeometry
$$
%$$
%\eqalign{
%& \nu_{01} - U \nu_{00}~=~ -6 \pi i~({\cal G}_{(1)T} +{\cal
%G}_{(2)TTU})
%\ , \quad  \nu_{10} - T \nu_{00}  ~=~ - 6 \pi i~({\cal G}_{(1)U} +
%{\cal G}_{(2)TUU}) \ , \cr  & \nu_{02} - 2 U \nu_{01} + U^2
%%\nu_{00}~=~
%-24 \pi i ~{\cal G}_{(2)TT}  \ , \quad
%\nu_{20} - 2 T \nu_{10} + T^2 \nu_{00} ~=~ -24 \pi i ~{\cal G}_{(2)UU}
%\ ,
%\cr  & \nu_{11} -  U \nu_{10} -  T \nu_{01}  + TU \nu_{00} ~=~
%-6 \pi i ~({\cal G}_{(1)} + 3 {\cal G}_{(2)TU}) \cr &
%(\nu_{12} -  2 U \nu_{11} + U^2 \nu_{10}) - T (\nu_{02} -  2 U
%%\nu_{01}
%+
%U^2 \nu_{00}) ~=~  - 72 \pi i ~{\cal G}_{(2)T}   \cr &
%(\nu_{21} -  2 T \nu_{11} + T^2 \nu_{01}) - U (\nu_{20} -  2 T
%%\nu_{10}
%+
%T^2 \nu_{00}) ~=~  - 72 \pi i ~{\cal G}_{(2)U}  \cr &
%(\nu_{22} -  2 U \nu_{21} + U^2 \nu_{20}) - 2 T (\nu_{12} -  2 U
%\nu_{11}
%+ U^2 \nu_{10}) + \cr & \qquad \qquad \qquad \qquad \qquad \qquad
%T^2 (\nu_{02} -  2 U \nu_{01} + U^2 \nu_{00}) ~=~
%216 \pi i ~{\cal G}_{(2)}   \ . } \eqn\Gtworelns
%$$

One can prove these relations by first differentiating both sides
sufficiently often with respect to $T$ until the left-hand side
can be simplified using \diffidents\  combined with the differential
equations satisfied by the $\mu_{jk}\lb 4$, while the right-hand side
is simplified using \derivpreps.  This process is then
repeated for the $U$-derivatives of the \Gtworelns.  The success
of this procedure critically depends on the proper form of
the \GPFsource\ and provides a significant number of
non-trivial tests upon the form  of \GPFsource.

\subsec{Periods of a Five-Fold?}

Eq.\ \newspecialgeometry\ is a direct analog of the classic
special geometry relation \FfromPF, and reflects
how the periods of the suspected five-fold would assemble
into the prepotential. It thus appears as a good
starting point for unraveling the analog of special
geometry in eight dimensions.

In this context, it is instructive to go one step  further and try to
infer how \otherform\ and  the prepotential $\Fn4(T,U)$ could arise
from a PF system of a $5$-fold and a corresponding  prepotential
$\FSTU(S,T,U)$, respectively. Recall that for $\Fn2(T,U)$  and the
$3$-fold the periods are $\pi_0, S \pi_0, T \pi_0, U \pi_0$  and ${\cal
F}_S \pi_0$, ${\cal F}_T \pi_0$, ${\cal F}_U \pi_0$,  ${\cal F}_0
\pi_0$, and in the $S \to \infty$ limit  $\pi_0, T \pi_0, U \pi_0$ and
${\cal F}_S \pi_0$ become the  periods of the $K3$ fiber, while the
finite parts of  $ {\cal F}_T \pi_0, {\cal F}_U \pi_0$ and ${\cal F}_0
\pi_0$  satisfy the $K3$ PF system with sources, and give rise to
$\Fn2(T,U)$ and its first derivatives.

Based upon this, and remembering the structure \homsols\ of the
homogenous solutions, we conjecture (in line with the findings of \LSW)
that the $5$-fold is the hyper-K\"ahler $4$-fold Sym$^2(K3)$ fibered
over a $\IP^1$ base.  As mentioned above, the periods of the fiber are
$T^j U^k{\varpi_0}^2$, $j,k =0,1,2$,  and these arise in the $5$-fold
as the $S\to \infty$  limit of $\pi_0, S \pi_0, T \pi_0, U \pi_0$ and
$\FSTU\lb4_{S} \pi_0,\FSTU\lb4_{ST} \pi_0,\FSTU\lb4_{SU} \pi_0$,
$\FSTU\lb4_{STT} \pi_0, \FSTU\lb4_{STU} \pi_0, \FSTU\lb4_{SUU} \pi_0$.
Thus only the fiber periods that are linear in $T$ and $U$ are realized
directly. {}From \Gtworelns\ it appears that only  the derivatives
$\FSTU\lb4_{TTU}$, $\FSTU\lb4_{TUU}$, and  by extension
$\FSTU\lb4_{STT} \pi_0, \FSTU\lb4_{STU} \pi_0$ and $\FSTU\lb4_{SUU}
\pi_0$ will actually appear directly  as $5$-fold periods.  Moreover,
as with $\FSTU_0\lb2$,  lower order derivatives of $\FSTU\lb4$  will
appear in the periods as combinations like  $\FSTU\lb4_{TT} + {1 \over
2}U(\FSTU\lb2_{T} +\FSTU\lb4_{TTU})$.  The proper combinations are
inferred from how the source  equations arise in the $S \to \infty$
limit of the $3$-fold,  and based upon this we expect that the
combinations that would arise  from a $5$-fold will be those of the
form $\mu_{jk}\lb4 - T^j U^k  \mu_{00}\lb4$.

We could not explicitly verify this conjecture, simply because there is
no known algebraic representation of Sym$^2(K3)$, and even less, of the
relevant $\IP^1$ fibration of it. An algebraic or toric representation
would however be necessary for obtaining the Picard-Fuchs system. The
closest one seems to be able to get at, is the beautiful construction
of Beauville and Donagi \BD, which leads to the periods and
Picard-Fuchs equations of the holomorphic $(2,0)$-form of Sym$^2(K3)$.
Unfortunately, there does not seem to be any simple way to obtain from
this the periods of the $(5,0)$-form of the $\IP^1$ fibration.

In the absence of such explicit algebraic representations, we can thus
far only conclude that our results provide further evidence for the
conjectured five-fold, augmenting the findings of ref.\ \LSW.
Summarizing, our main results supporting this structure are: a) the
form \homsols\ of the homogenous solutions, which corresponds to a
fibration of Sym$^2(K3)$, and b) the writing \newspecialgeometry\ of
the prepotential $\Fn4(T,U)$ in terms of the inhomogenous solutions of
the PF equations.

%%%%%%%%%%%%%%%%%%%%%%%%%%%%%%%%%%%%%%%%%%%%%
\chapter{Some remarks on curve counting}
%%%%%%%%%%%%%%%%%%%%%%%%%%%%%%%%%%%%%%%%%%%%%

In the compactification to four dimensions, sending $S\to\infty$
corresponds to the large base space limit of the $K3$ fibration.
Therefore, the coefficients $c\lb2$ of $\ellgen_{-2}(q)$ in \gentwo\
must correspond to counting certain ``rational curves'' in the $K3$
fiber. However, it is known that other $K3$ fibrations lead to
different counting functions, see, for example, \gm.
Moreover, a generic $K3$ has no rational curves at all. Counting
rational curves in $K3$ thus depends upon how one broadens the concept.
By considering $\ellgen_{-2}(q)=E_4E_6/\eta^{24}$ we count the
$2$-cycles in $K3$ that become rational curves in our particular choice
of fibration over $\IP^1$.

The most canonical way to count rational curves in $K3$ was presented
in \YZ, where one counts certain singular curves that are holomorphic
in a given, fixed complex structure; the relevant counting function in
this instance is simply given by $\eta^{-24}$. As was shown in \BL,
this can be obtained by the trivial fibration $K3\times \IP^1$, where
$\IP^1$ corresponds to the twistor family of complex structures in the
hyper-K\"ahler $K3$. This reasoning does not involve mirror symmetry,
and indeed $K3\otimes \IP^1$ is not even a Calabi-Yau space.

Our point is that it is in eight dimensions where one can compute the
counting function $\eta^{-24}$ via mirror symmetry. More precisely, in
our computation the counting function was
$\ellgen_{-4}(q)={E_4}^2/\eta^{24}$ \genfour, and the difference as
compared to four dimensions is that the $E_4$'s can be removed by
incorporating the $E_8\times E_8$ Wilson lines $\vec V$ in the
prepotential. That is, as mentioned in \LS, extending the sum over the
$E_8\times E_8$ lattice one can write
$$
\Fn4(T,U,\vec V)\ \sim\
\sum_{{(k,l,\vec r)>0\atop \vec r\in \Lambda_{E_8\times E_8}}}
\tilde c\lb4(kl-\vec r^{\,2}/2)\
\Li_5\big[e^{2\pi i(kT+l U+\vec r\cdot\vec V)}\big]\ ,
\eqn\totalG
$$
where
$$
\eta(q)^{-24}\ \equiv\  {1\over q}\prod_{l\geq1} (1-q^l)^{-24}
\ =:\ \sum_{n\geq -1} \tilde c\lb4(n)q^n
\eqn\thetageneratingfunction
$$
is exactly the counting function of \doubref\YZ\BL. This function  is
known to count 1/2-BPS states in $K3$ compactifications of the IIA
theory  \BSV.  Here we find that it also counts 1/2-BPS states in
$F$-theory on $K3$, in line with the arguments in
\BK\ for the heterotic string in eight dimensions.

Thus, what we have been arguing in this paper is, essentially,  how to
determine this counting function via the mirror map.\foot{To do this
completely would require incorporating the 16 Wilson line moduli in the
differential equations.} While on the one hand $K3\times \IP^1$ is not
a Calabi-Yau space, and on the other, non-trivial $K3$ fibrations over
$\IP^1$ do not lead to $\eta^{-24}$, it appears that the appropriate
geometry to obtain \thetageneratingfunction\ from mirror symmetry is a
fibration of Sym$^2(K3)$ over $\IP^1$.

%%%%%%%%%%%%%%%%%%%%%%% %%%%%%%%%%%%%%%%%%%%%%%%%%%%%%%%
\ack
%%%%%%%%%%%%%%%%%%%%%%% %%%%%%%%%%%%%%%%%%%%%%%%%%%%%%%%

We would like to thank
R.\ Dijkgraaf,
R.\ Donagi,
C.\ Doran,
A.\ Klemm,
P.\ Mayr,
D.\ Morrison,
N.\ Nekrassov,
C.\ Vafa,
and E.\ Zaslow
for discussions. W.L.\ also thanks the Harvard Physics Department
for kind hospitality.

%%%%%%%%%%%%%%%%%%%%%%%%%%%%%%%%%%%%%%%%%%%%%
\appendix{A}{Formal extension to $n=6$}
%%%%%%%%%%%%%%%%%%%%%%%%%%%%%%%%%%%%%%%%%%%%%

The mathematical structure of the prepotentials \prep\ can be
considered for any value of $n$, and for any modular form
$\ellgen_{-n}(q)$ of weight $-n$. From the physical point of view the
generalization appears to be purely formal. On the ``heterotic side''
we would be need to start in $2n+2$ dimensions, and consider a toroidal
compactification to give an amplitude (Tr $(F^n)$) in $2n$-dimensions.
Of course, there are no such superstrings for $n>4$. However, the
situation is reminiscent of anomaly cancellation \anom, and is indeed
related to it: the mathematical mechanism is very general, being just
based on modular properties of the elliptic genus, and works in
``string theories'' in any dimension, no matter how pathological their
physical meaning.

We demonstrate here that the prepotential \prep\  makes
formally sense for $n=6$, even though there is no known consistent
string theory whose amplitudes it would describe. Just from modular
properties we must have that $\cA_{-6}(q)=E_6/\eta^{24}$ and so the
relevant ``one-loop amplitudes'' are of the form:
$$
\eqalign{
\Delta_{\F_T^6}&={(U-\ov U)^3\o (T-\ov T)^3}
\int {d^2 \tau \o \tau_2} \sum_{(p_L,p_R)}\ \ov p_R^6\
q^{\h |p_L|^2} \ov q^{\h |p_R|^2}\ {\ov E_6\o \ov \eta^{24}}\ ,\cr
\Delta_{\F^3_T \F_U^3}\!&=
\!\int\!\! {d^2 \tau \o \tau_2}\!\!
\sum_{(p_L,p_R)}\!\lf[|p_R|^6\!-
{9\o 2\pi \tau_2} |p_R|^4\!+\!{9\o 2\pi^2 \tau_2^2} |p_R|^2\!
-{3\o 4\pi^3\tau_2^3}\ri]q^{\h |p_L|^2}
\ov q^{\h |p_R|^2}
{\bar E_6\o \ov \eta^{24}}\cr}
\eqn\intFV
$$
and similar expressions for $\Delta_{\F^5_T \F_U},\
\Delta_{\F^4_T\F_U^2}$.  These couplings integrate to one and
the same holomorphic prepotential $\Fn6(T,U)$, given by \prep\ for
$n=6$. Explicitly:
$$
\eqalign{
\Delta_{\F_T^6}&=-32\pi i\lf(\p_T+{4\o T-\ov T}\ri)\lf(\p_T+
{2\o T-\ov T}\ri)\p_T\cr
&\ \ \ \ \times\lf(\p_T-{2\o T-\ov T}\ri)
\lf(\p_T-{4\o T-\ov T}\ri)\lf(\p_T-{6\o T-\ov T}\ri)\Fn6(T,U)\cr
&+32\pi i{(U-\ov U)^6 \o (T-\ov T)^6}
\lf(\p_\Uc-{4\o U-\ov U}\ri)\lf(\p_\Uc-{2\o U-\ov U}\ri)\p_\Uc\cr
&\ \ \ \ \times\lf(\p_\Uc+{2\o U-\ov U}\ri)
\lf(\p_\Uc+{4\o U-\ov U}\ri)\lf(\p_\Uc+{6\o U-\ov U}\ri)\ov
\Fn6(T,U)\cr
\Delta_{\F^3_T \F_U^3}
&=- 32\pi i
\lf(\p_U-{2\o U-\ov U}\ri)\lf(\p_U-{4\o U-\ov U}\ri)
\lf(\p_U-{6\o U-\ov U}\ri)\cr
&\ \ \ \ \times
\lf(\p_T-{2\o T-\ov T}\ri)\lf(\p_T-{4\o T-\ov T}\ri)\lf(\p_T-{6\o T-\ov
T}\ri)
\Fn6(T,U)+hc.\ .\cr}\eqn\DeltaFV
$$
The correction $\Delta_{\F_T^6}$ represents a function of weights
$(w_T,w_U)=(6,-6)$ and $(w_{\ov T},w_{\ov U})=(0,0)$, respectively.
While it is not fully harmonic, a holomorphic, covariant quantity may
be obtained via an additional $T$--modulus insertion, by considering
$$
\Fn6_{TTTTTTT}=
{i\o 16} {(U-\ov U)^3\o (T-\ov T)^4}
\int d^2 \tau \sum_{(p_L,p_R)}p_L\ov p_R^7\
q^{\h |p_L|^2}\ov q^{\h |p_R|^2}\ {\bar E_6\o \ov \eta^{24}}\ .
$$
It is a non-trivial feature that this integral indeed yields a
holomorphic covariant quantity:
$$
\Fn6_{TTTTTTT}\ =\ \prod_{k=-3}^3\lf(\p_T-{2k\o T-\ov T}\ri)\Fn6\
=\ {E_4(T)^2E_6(U)  \o [J(T)-J(U)]\eta^{24}(U)}\ ,
\eqn\NICE
$$
and similarly for the other couplings. For example,
$$
\Fn6_{TTTTUUU}\
=\ {1\o 2\pi i}{\p_T}\log\big[J(T)-J(U)\big] + {1\o 2\pi i}{\p_T}
\ln\Psi_0(T,U)\ ,
\eqn\link
$$
where
$$
\eqalign{
\Psi_0(T,U)=q_T \prod_{(k,l)>0}\lf(1-{q_T}^k {q_U}^l\ri)^{d(kl)}\ .}
\eqn\PSI
$$
The cusp form $\Psi_0$ stays finite everywhere in the moduli space,
i.e.,
$d(-1)=0= d(0)$, and the exponents are generated by
%$$
%\sum_{n>0} d(n)q^n\ =\ {5 \o 72} {E_2^3E_6\o \eta^{24}}   +{5\o 24}
%{E_2^2E_4^2\o \eta^{24}}+{3\o 8}{E_2E_4E_6\o \eta^{24}}
%-{11\o 36}{E_6^2\o \eta^{24}}-{25\o 72}{E_4^3\o \eta^{24}}\ .
%%  &=-168480q-16035840q^2-581878080q^3+\ldots\ .
%\eqn\dcoeff
%$$
$
\sum_{n>0} d(n)q^n = \big({5 \o 72} {E_2^3E_6}   +{5\o 24}
{E_2^2E_4^2}+{3\o 8}{E_2E_4E_6 }
-{11\o 36}{E_6^2 }-{25\o 72}{E_4^3 }\big)/\eta^{24}
$.

Moreover, as we have indicated above, the relationship between  the
functions $\Fn{n}(T,U)$ and PF systems with sources also appears to
generalize in a natural manner.   As discussed above, $\partial_T^7
{}~\Fn6(T,U)$ is given by \derivpreps. Following the algorithm outlined
above we find the  function:
$$
\mu_{00}\lb6   ~=~  2 \pi i(\Fn6_{TTTUUU}~+~ 5 \Fn4_{TTUU}
{}~+~ 9 \Fn2_{TU})  ~-~ 5 log(w_1 -w_2) \ .
$$
It satisfies \generalform\ for $n=6$ in which the homogenous part,
${\cL}_{w_a}\ot3$ is given by \threeDOs, and the source part by:
$$
\eqalign{
\cM\lb{6}_1 ~:=~ -~
& {20  w_2 \over (w_1 - w_2)}~  \bigg[ (1-w_1)\big( {\cL}_{w_1}\ot2 -
\coeff{5}{48} \twa - \coeff{5}{144} \big) ~-~ \cr
& \qquad  \qquad (1-w_2)\big( {\cL}_{w_2}\ot2 -
\coeff{5}{48} \twb - \coeff{5}{144} \big) ~-~ \cr
& \qquad  \qquad w_1 (1-w_2) {d \over d w_2} \big({\cL}_{w_1}\ot1 -
\coeff{5}{72} \big) ~+~ \cr
& \qquad \qquad w_1 (1-w_1) {d \over d w_1}
\big( {\cL}_{w_2}\ot1 - \coeff{5}{72} \big)\bigg] \cr
{} +~ & 5~ \big( w_2 {\cL}_{w_2}\ot1 + \coeff{1}{6} (1-w_2) \twb -
\coeff{1}{12} w_2 \big) \ .}
\eqn\HPFsource
$$
The structure of the homogenous equations is indeed
that of the PF equation of Sym$^3(K3)$.
\nobreak

%%%%%%%%%%%%%%%%%%%%%%%%%%%%%%%%%%%%%%%%%%%%%%%

\goodbreak
%\vfil\eject
\refout
\vfill
\eject
\end